\documentclass[journal=jctcce,manuscript=article]{achemso}
\usepackage{color}
\usepackage[version=3]{mhchem} 
\usepackage[T1]{fontenc}       
\usepackage[utf8x]{inputenc}

\author{Sudipta Kundu}
\affiliation{Centre for Condensed Matter Theory, Department of Physics, Indian Institute of Science, Bangalore 560012, India}
\author{Satadeep Bhattacharjee}
\affiliation{Indo-Korea Science and Technology Center, Bangalore 560065, India}
\author{Seung-Cheol Lee}
\affiliation{Indo-Korea Science and Technology Center, Bangalore 560065, India}
\author{Manish Jain}
\email{mjain@iisc.ac.in}
\affiliation{Centre for Condensed Matter Theory, Department of Physics, Indian Institute of Science, Bangalore 560012, India}

\title
{Population Analysis with Wannier Orbitals}

\begin{document}

\begin{abstract}
We formulate Wannier orbital overlap population and Wannier orbital Hamilton population to describe the contribution
of different orbitals to electron distribution and their interactions. These methods, which are analogous to the 
well known crystal orbital overlap population and crystal orbital Hamilton population, provide insight into the
distribution of electrons at various atom centres and their bonding nature. We apply this formalism in the context of
a plane-wave density functional theory calculation. This method provides a means to connect the non-local plane-wave basis
to a localised basis by projecting the wave functions from a plane-wave density functional theory calculation on to localized Wannier orbital basis.
The main advantage of this formulation is that the spilling factor is strictly zero for insulators and can systematically be made small
for metals. We use our proposed method to study and obtain bonding and electron localization insights in five different materials.
\end{abstract}


\section{Introduction}
Electrons localize differently
in different types of materials. In ionic materials electrons are localized at
atomic centres; in metals they are itenerant;  while they accumulate at bond centres in covalent materials.
Distribution of electrons and their bonding behavior are of paramount importance
to understand a material characteristics. Density functional theory (DFT) is well
known for studying ground state properties of matter \cite{dft1,dft2}. It has seen
widespread success in describing the ground state properties such as total energy,
electron density, structure, dynamical properties etc \cite{dft_eg1,dft_eg2}.

Plane wave based DFT codes \cite{pw1,pw2,vasp3,vasp4,qe,abinit,castep} are
widely used due to the flexibility, simplicity and accuracy of the plane-wave
basis set. The plane-wave basis is homogeneous and its completeness is controlled 
by a single parameter: the energy cutoff (which determines the number of plane-waves
included in the calculation). While plane-wave basis set offers accuracy and
simplicity, it is
unsuitable for chemical analysis owing to its non-local nature. On the other hand
localized basis sets offer a natural route to chemical analysis. However, systematic convergence
of ground state properties with these bases is much more difficult and the final results
can depend on the particular choice of basis.

Population analysis based on localized basis sets offers a chemically intuitive way to
move from wave functions to bond order description of bonding in materials. Several population
analysis schemes using basis set localized at atoms have been proposed. For example, Mulliken
population analysis is used to characterize the distribution of electrons in a molecule
or the bonding, anti-bonding and non-bonding nature of interaction for a pair of atoms\cite{mulliken,mulliken2}.
Mulliken population analysis is based on molecular orbitals which are constructed using atomic orbitals.
For extended systems, a useful and popular technique was introduced by Hughbanks
and Hoffmann within the framework of extended Huckel theory \cite{huckel1,huckel2}. Another set
of methods, crystal orbital
overlap population (COOP) and crystal orbital Hamilton population (COHP), are based on extending the existing
Mulliken population analysis to solid state systems \cite{coop1,coop2,coop3,cohp1,cohp2}. 
COOP is the solid-state analogue to the molecular bond order. 

Due to its intuitive nature, population analysis has become an important tool in DFT as well. Localized
atomic orbital basis based DFT code \texttt{SIESTA} \cite{siesta} has their own implementation of calculating COOP and COHP.
Another linear scaling DFT package \texttt{ONETEP} \cite{onetep} employs density derived electrostatic and chemical method to compute atomic
charges \cite{ddec}. Onetep uses nonorthogonal generalized Wannier functions as basis set. 
Recently, there has been a lot of work to connect the plane-wave basis set based DFT codes to
localized basis sets by projecting the eigenfunctions of a plane-wave calculation
into Hilbert space spanned by the local basis. 
There have been several approaches to project the electronic eigenfunctions onto local basis or 
onto orbitals that are constructed from pseudopotential used in plane-wave calculation \cite{method1,method2,spill,method3,method4,method5,method6}.
One such code, \texttt{LOBSTER}, projects PAW (projector augmented wave) wave functions obtained from DFT calculation using \texttt{VASP}, \texttt{Quantum Espresso}
or \texttt{ABINIT} into a LCAO (linear combination of atomic orbitals) basis to calculate projected COOP, COHP and  density of states (DOS) \cite{lobster1,lobster2}.

Here we propose an alternative population analysis method using Wannier orbitals as basis set.
Population analysis using Wannier functions has been proposed previously by Bhattacharjee et al. \cite{woop}
In their method, they proposed to project Wannier functions of occupied subspace to Wannier functions of extended subspace.
Our formulation is different as we use only one set of Wannier functions to expand plane-wave eigenfunctions.
Amongst different ways of constructing Wannier functions \cite{RevModPhys.84.1419,wan3,wan2,scdm,scdm2,diff_wan} in periodic systems, we choose the maximally 
localized Wannier functions (MLWF) \cite{RevModPhys.84.1419,wan3,wan2,wan1} 
as the basis functions. MLWF are localized functions which
can be readily constructed from the Bloch eigenfunctions calculated within a plane-wave DFT code. These
local basis functions are orthonormal and provide a good basis for writing a compact
Hamiltonian.  As Wannier functions are obtained by performing unitary transformations on Bloch functions,
there is no loss of information.  The Wannier orbitals are then employed to calculate overlap population
and Hamilton population. Wannier orbital overlap population (WOOP) describes the distribution of
electrons among different orbitals. Wannier orbital Hamilton population (WOHP)
describes the interaction energy between different orbitals. 
It is to be noted that our proposed formalism can be used with any kind of Wannier functions \cite{scdm,scdm2,diff_wan} and
is not limited to MLWFs. In order to assess the proposed formulation, 
we have calculated WOOP and WOHP for five different materials. We use diamond to benchmark our implementation. To show
the utility and validity of the proposed method, we 
extend our study to other systems: Gallium Arsenide (GaAs), titanium (Ti), carbon
nanotube (CNT) and La$_2$NiMnO$_6$ (LNMO) double perovskite. 

\section{Theoretical formalism}
Electronic structure calculations using a plane-wave basis are carried out under periodic boundary conditions. As
the effective potential calculated within DFT in a periodic solid is periodic at the unit cell level, the effective one particle Hamiltonian 
commutes with the lattice translation operator. This allows one to write the solution eigenfunctions of this effective
one particle Hamiltonian in a Bloch form. The $\mu$th Wannier function, $w_\mu(\textbf{r},\textbf{R})$, located in the unit cell at
$\textbf{R}$ with respect to origin, is defined using a Fourier transform of the
Bloch functions, $\psi_{m\textbf{k}}(\textbf{r})$, \cite{RevModPhys.84.1419} as:
\begin{equation*}
w_\mu(\textbf{r},\textbf{R}) = \frac{V}{(2\pi)^3}\int_{\text{BZ}} d\textbf{k} \; e^{-i\textbf{k}.\textbf{R}}\sum_{m}U_{\mu m}^{\textbf{k}}\psi_{m\textbf{k}}(\textbf{r})
\end{equation*}
where $U_{\mu m}^{\textbf{k}}$ are unitary matrices which combine
the Bloch functions $\psi_{m\textbf{k}}(\textbf{r})$, $\textbf{k}$ are wavevectors in the first Brillouin zone and $m$
is the band index. Owing to 
due to the 'guage freedom' of Bloch eigenfunctions \cite{RevModPhys.84.1419}, within the maximally localized Wannier function formulation, $U_{\mu m}^{\textbf{k}}$ are chosen
in a way that minimizes the spread of the Wannier functions.
By construction, this set of MLWF is
orthonormal. 

Having constructed the MLWF, the eigenfunctions obtained within DFT at any $\textbf{k}$ point can be expanded in terms of
the Wannier functions ($\lvert w_{\mu\textbf{R}}\rangle$) as:
\begin{equation*}
\lvert\psi_{j\textbf{k}}\rangle\approx\sum_{\mu\textbf{R}}C_{\mu\textbf{R},j}(\textbf{k})\lvert w_{\mu\textbf{R}}\rangle \equiv \lvert w_{j\textbf{k}}\rangle
\end{equation*}
where $C_{\mu\textbf{R},j}$ is the expansion coefficient of $j$th Bloch wave function with respect to the $\mu$th
Wannier orbital basis located in the unit cell at $\textbf{R}$ and $\lvert w_{j\textbf{k}}\rangle$ is the best approximation to $\lvert\psi_{j\textbf{k}}\rangle$ that
one obtains upon expansion in the Wannier basis.
$C_{\mu\textbf{R},j}$ is defined as:
\begin{equation*}
C_{\mu\textbf{R},j} = \frac{V}{(2\pi)^3}\int_{\text{BZ}} d\textbf{k} \; e^{i\textbf{k}.\textbf{R}} \tilde{U}^{\textbf{k}}_{\mu j}
\end{equation*}
where the matrix, $\tilde{U}$, is the inverse of $U$.

Using the above definition of expansion coefficients, one can write WOOP (analogous to COOP) as:
\begin{equation*}
WOOP_{\mu\textbf{R},\nu\textbf{R'}}(E) = S_{\mu\textbf{R},\nu\textbf{R'}}\sum_{j,\textbf{k}}C^*_{\mu\textbf{R},j}(\textbf{k})C_{\nu\textbf{R'},j}(\textbf{k})\delta(\epsilon_j(\textbf{k})-E) \\
\end{equation*}
where $S_{\mu\textbf{R},\nu\textbf{R'}}$ is the overlap between $\mu$th Wannier function at $\textbf{R}$ with $\nu$th Wannier function at $\textbf{R'}$. In
the above equation and all the subsequent equations, the integral over \textbf{k} in the entire Brillouin zone has been replaced by a discrete sum over all the \textbf{k} as is 
routinely done in all calculations. Due to the
orthogonality of Wannier functions, $S_{\mu\textbf{R},\nu\textbf{R'}} = \delta_{\mu\textbf{R},\nu\textbf{R'}}$, the expression for WOOP reduces to:
\begin{equation*}
WOOP_{\mu\textbf{R}}(E) = \sum_{j,\textbf{k}}C^*_{\mu\textbf{R},j}(\textbf{k})C_{\mu\textbf{R},j}(\textbf{k})\delta(\epsilon_j(\textbf{k})-E) 
\end{equation*}
From the above equation it is clear that WOOP for an orbital can be interpreted as the partial density of states associated with that orbital. 
Integrating WOOP upto the Fermi level for an orbital gives the number of electrons associated with it.
When summed over all orbitals, WOOP gives total density of states. It should be noted that this 
formulation of population analysis is conceptually different from the usual crystal overlap formulation. In
this formulation of WOOP, one constructs atomic-orbital-like Wannier functions from the Bloch functions
and calculates orbital overlap population with those functions. On the other hand, COOP employs atomic orbital basis which
requires inclusion of a large number of orbitals to accurately represent the system.
Furthermore, as the atomic orbitals are not orthogonal, COOP has off-diagonal contributions while
they are strictly zero for WOOP.

Extending the analogy from COHP, WOHP can be written as:
\begin{equation*}
WOHP_{\mu\textbf{R},\nu\textbf{R'}}(E) = -H_{\mu\textbf{R},\nu\textbf{R'}}\sum_{j,\textbf{k}}C^*_{\mu\textbf{R},j}(\textbf{k})C_{\nu\textbf{R'},j}(\textbf{k})\delta(\epsilon_j(\textbf{k})-E)
\end{equation*}
where $H_{\mu\textbf{R},\nu\textbf{R'}}$ is hopping matrix element between $\mu$ Wannier orbital at $\textbf{R}$
and $\nu$ Wannier orbital at $\textbf{R'}$. WOHP is energy weighted WOOP for the on-site term ($\mu=\nu$). 
For $\mu\neq\nu$ (off-site term), WOHP provides a way to evaluate the strength of interaction between those two orbitals and 
compare interactions between different orbitals. Although WOHP can be complex number in general, we find that the imaginary part
is very small compared to the real part and hence we analyse the real part to interpret the chemical character \cite{jcc1}.
Positive WOHP signifies bonding states while anti-bonding states are represented by
negative WOHP. 

The main advantage of using Wannier functions as a basis set can be understood by their ability to represent the eigenfunctions.
In order to quantify this better, one can use the spilling factor\cite{spill} defined as:
\begin{align*}
S&=\frac{1}{N_{\textbf{k}}}\frac{1}{N_b}\sum_{j,\textbf{k}}\langle\psi_{j\textbf{k}}\lvert(1-P_{\textbf{k}})\lvert\psi_{j\textbf{k}}\rangle\\
&=\frac{1}{N_{\textbf{k}}}\frac{1}{N_b}\sum_{j,\textbf{k}}[1-\sum_{\mu}\lvert\langle\psi_{j\textbf{k}}\lvert w_{\mu\textbf{k}}\rangle\lvert^2]
\end{align*}
where, $P$ is a projection operator given by $P_{\textbf{k}}=\sum_{\mu}\lvert w_{\mu\textbf{k}}\rangle\langle w_{\mu\textbf{k}}\lvert$, $N_b$ is the
number of Bloch bands used and $N_\textbf{k}$ is the number of $\textbf{k}$-points in the Brillouin zone. 
The spilling factor has a lower limit of 0 and an upper limit of 1. $S = 0$ implies
that the wave functions projected on local basis span the same space as the original
eigenfunctions. In general, lower the value of S, better is the local basis. In
contrast, $S = 1$ signifies that the projected wave functions are orthogonal to 
to the original eigenfunctions.

While constructing atomic orbital like Wannier functions, two types of cases
are encountered: a) Wannier functions constructed from 
isolated set of bands, b) Wannier functions constructed from entangled bands. In the first case, the number of Bloch functions is
the same as the number of Wannier orbitals. As a result, the $U$ matrix 
combines and rotates the Bloch functions to get required Wannier orbitals. In this case, the spilling factor,
S, can be written as:
\begin{equation*}
S=\frac{1}{N_{\textbf{k}}}\frac{1}{N_b}\sum_{j,\textbf{k}}[1-\sum_{\mu}\tilde{U}^{\textbf{k}}_{\mu j}\tilde{U}^{*\textbf{k}}_{j\mu}]
\end{equation*}
Due to unitarity of $U$ matrices and hence of $\tilde{U}$ matrices, $S=0$. Thus, for isolated bands (for
example insulators), the spilling is strictly zero.
As a result, all the information obtained from DFT eigenfunctions is retained in the Wannier orbitals and is used to interpret chemical properties. 
One consequence of this is that the WOOP summed over all the orbitals and integrated up to the Fermi energy would give the correct number of electrons.
The second case arises when we have a metal or when the bands within the energy range we are
interested in are connected with bands of energy outside the range. In these cases a disentanglement 
of bands has to be performed prior to wannierisation. A method for disentanglement proposed by Souza et al.\cite{wan2} is most commonly used
to obtain the correct subspace of Bloch functions. Wannier
orbitals are constructed using within this space of Bloch functions. 
For these systems, the Bloch energy states required for correct representation of the Wannier
function can span a wide energy range and as a result spilling factor can be non-zero.
It can, however, be made smaller by simply including more bands
in the disentanglement procedure to obtain the correct subspace.

\section{Application to different material systems}
We have performed WOOP and WOHP calculations on five materials. We have chosen these materials to highlight the
usefulness of this approach. All DFT calculations for these examples are
performed using the \texttt{Quantum Espresso}\cite{qe} package.
We use ultrasoft pseudopotential \cite{ultrasoft} for carbon atom in example diamond and CNT.
We adopt PAW pseudpotential \cite{paw} for GaAs and LNMO and ONCV 
pseudopotential \cite{oncv1,oncv2} for titanium. Usage of different pseudopotentials among 
different examples is possible because construction of Wannier function does not depend on
the psuedopotential directly and hence provides flexibility in choice of pseudopotential to calculate WOOP and WOHP.
The exchange-correlation functional is approximated using generalized gradient approximation parametrized by Perdew Burke and Ernzerhof (PBE)\cite{pbe}.
We use a plane-wave cutoff energy of 50 Ry to construct the plane-wave
basis functions used in the expansion of the DFT eigenstates for diamond.
The plane-wave cutoff is 35 Ry, 100 Ry, 30 Ry and 60 Ry for GaAs, Ti, CNT and LNMO respectively.
We use a $12\times12\times12$ Monkhorst-Pack k-point sampling \cite{kpt}of the Brillouin
zone to converge the charge density of diamond. The Brillouin zone is sampled with $4\times4\times4$, $9\times9\times5$, $1\times1\times30$, $8\times8\times8$ k-grid for GaAs, Ti, CNT
and LNMO respectively for DFT calculations. For WOOP and WOHP calculations, we adopt a finer
k-point sample and use $50\times50\times50$, $35\times35\times35$, $36\times36\times20$,
$1\times1\times60$, $20\times20\times20$  k-grid for diamond, GaAs, Ti, CNT and LNMO respectively.
The \texttt{Wannier90}\cite{wan1} package is used for constructing MLWFs
and a modified version of the same package is used for the calculation of WOOP and WOHP of the systems.

\subsection{Diamond: A benchmark calculation}

\begin{figure}[t]
\begin{center}
\includegraphics[scale=0.5]{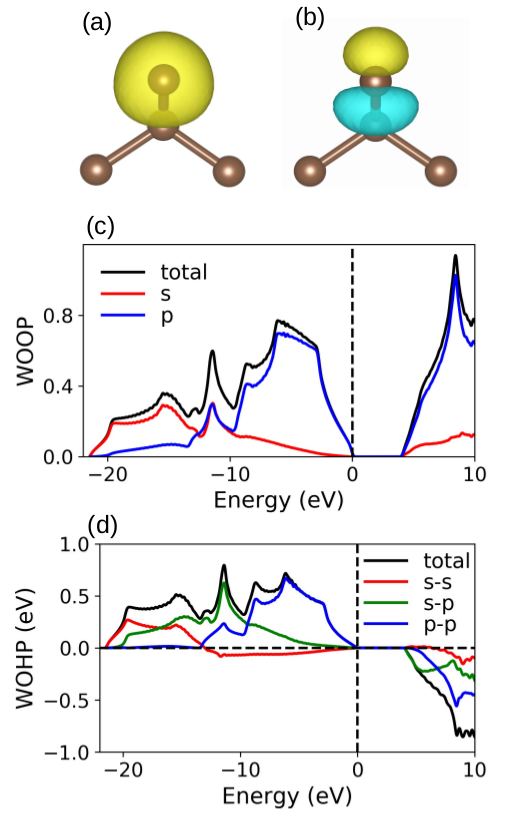}
\caption{(a) and (b) shows the constructed $s-$ and $p-$like Wannier functions at one of the C atom.
(c) depicts WOOP of diamond. The black, red and blue line shows the total WOOP and contribution due to $s-$ and $p-$type
orbitals respectively. (d) is total WOHP of diamond and shows contribution of interaction between different orbitals.
The valence band maximum is set to zero in (c) and (d).}
\end{center}
\end{figure}
Diamond consists of two inter-penetrating face centred cubic lattices displaced
by one-quarter along the body diagonal. In diamond, a C
atom is bonded with 4 others. The bonds are formed by hybridizing 2$s$ and 2$p$ atomic orbitals of C atoms into  $sp^3$ hybrid orbitals.
For calculation of WOOP
and WOHP we construct atomic orbital like Wannier functions at the carbon atom sites. The
orbitals look like $s$ and $p$ atomic orbitals, though the $p-$like orbitals do not have exactly the same shape as
the atomic $p$ orbital. Fig. 1(a) and (b) show the constructed $s$ and $p-$type orbitals
respectively.
In fig. 1(c) we show total WOOP (black) and contribution from
$s-$like orbital (red) and three $p-$like orbitals (blue). While the $s-$like
orbital has more contribution in low-lying energy bands, band edges are mostly
composed of $p-$orbitals. As WOOP summed over all the orbitals gives total density of states (DOS), we compare our results 
with Ref.[\citenum{jcc1}] and our results are in good agreement. We estimate the total number of
electrons to be 8 and hence spilling factor is zero. We also calculate
electrons associated with each orbitals. Each of the $s-$like orbital has 1.35 electrons while
each of the $p-$like orbitals holds 0.883 electrons. Ideally we expect the electrons to be equally distributed and
the number of electrons in each orbital to be 1. This is because the orbitals constructed via Wannier
functions are not exact atomic orbitals. In order to check this aspect, we construct $sp^3$-like orbitals
instead of pure atomic-like orbitals and calculate the electron numbers. We orient the $sp^3$-like orbitals
along the bond directions. We find that each of the $sp^3$-like orbitals holds 1 electron which is in 
line with our expectation.
We calculate WOHP and analyse the interactions among $s-$ and $p-$like orbitals of two C atoms. Fig. 1(d)
shows WOHP as a function of energy and compares the strength of interaction
between different type of orbitals. The $p-p$ interaction is stronger near valence band edge while $s-s$
interaction dominates deep inside the valence bands. In the intermediate energy region,
$\sim$ 9.5 -- 13.0 eV below the valence band maximum (VBM), $s-p$ interactions dominate. 
We set the VBM to zero
in fig. 1(c) and 1(d). Up to the VBM,
WOHP is positive signifying that the interactions are bonding. The higher states
are anti-bonding and hence addition of electrons will decrease the strength of the bonds.

\subsection{GaAs: A small band gap semiconductor}

\begin{figure}[h]
\begin{center}
\includegraphics[scale=0.5]{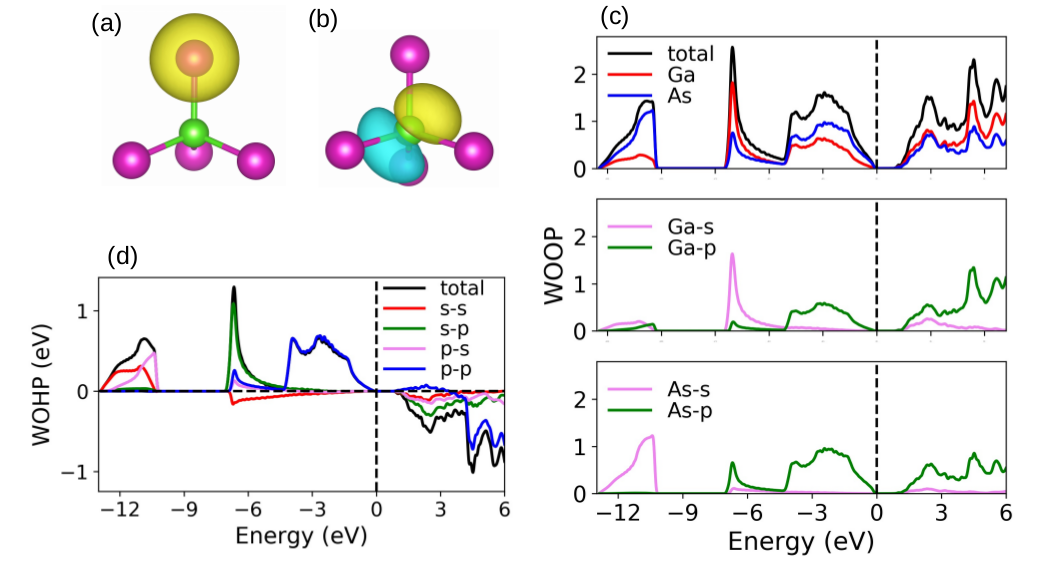}
\caption{(a) $s-$like orbital on Ga atom (purple) and (b) $p-$like orbital on As atom (green).
(c) Total WOOP and also contribution from $s$ and $p-$like orbitals of Ga and As atoms.
(d) Total WOHP and orbital resolved WOHP of GaAs. The first orbital of label corresponds to Ga atom and the 
second one corresponds to As atom. VBM is set to zero.}
\end{center}
\end{figure}
Gallium arsenide is a semiconductor with zinc blend structure. In GaAs, each Ga atom is bonded to
four As atoms and vice-versa. The 4$s$ and 4$p$ orbitals of both the Ga and As atoms contribute to the states near the band edges.
Using Wannier90, $s-$ and
$p-$like MLWFs on Ga and As atoms are constructed (fig. 2(a) and (b)).
We plot the total WOOP and the contribution of $s-$ and $p-$orbitals of Ga and As atom in fig. 2(c).
We find that near band
edges the major contribution is mostly from $p$ orbitals of both Ga and As atoms.
The peak near 6 eV below the VBM is primarily due to $s$ orbital of Ga and
the deep valence states are mostly comprised of $s$ orbital of As.
GaAs has ionic character due to the difference of electronegativity of Ga and As. As is more electronegative
and hence, electrons from Ga transfer to As. We calculate the number of electrons at the 
Ga and As sites to be 3.2 and 4.8 respectively. As GaAs is a small band-gap semiconductor and the MLWFs are constructed
from an isolated set of bands, spilling factor is zero.
Fig. 2(d) shows the orbital resolved WOHP of GaAs. 
The interaction between $p$ orbitals of Ga and As is predominant near valence band edge
consistent with the WOOP calculation. The intermediate region, around 6 eV below the VBM,
is dominated by the interaction between $s$ orbital of Ga and $p$ orbital of As atom whereas the deep states
have interaction between both $s$ and $p-$orbitals of Ga atom and $s$ orbital of As.
Here again as in the case of diamond, the interactions are bonding type upto the VBM and
conduction band states have antibonding character.

\subsection{Titanium: A metal}

\begin{figure}[h]
\begin{center}
\includegraphics[scale=0.5]{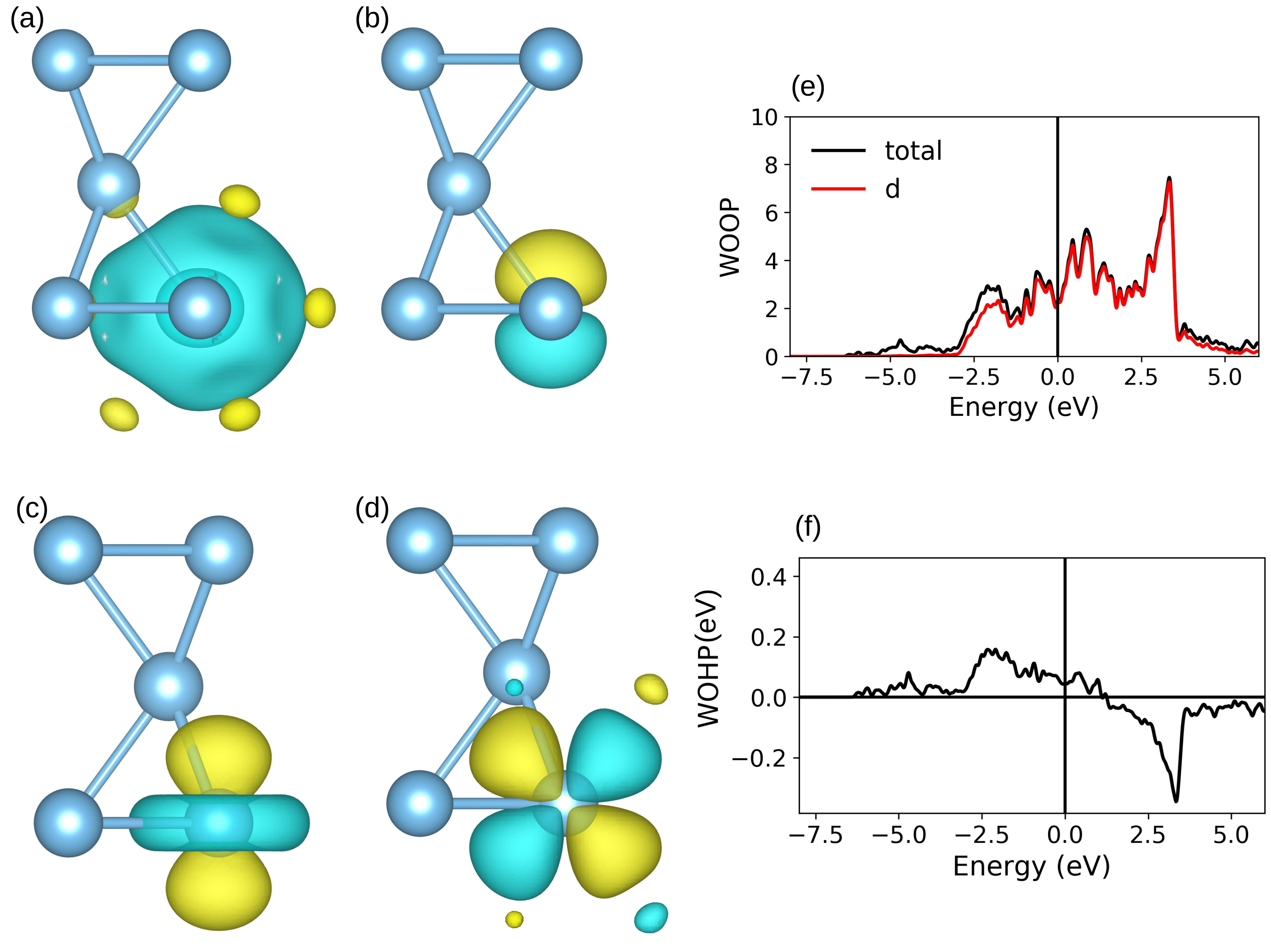}
\caption{(a) $s-$ and (b) $p-$like orbitals on Ti atom. (c),(d) are two $d-$like
orbitals. (e) depicts total WOOP (black). The contribution from $d-$like orbitals is shown in red. (f) shows the WOHP. Fermi level is set to zero.}
\end{center}
\end{figure}
Titanium is a transition metal. We explore the hexagonal closed pack structure of titanium.
The unit cell consists of two Ti atoms with electronic configuration [Ne]3s$^2$3p$^6$3d$^2$4s$^2$.
We are primarily interested in states near Fermi level.
It is expected that they will arise primarily from 3$d$ and 4$s$ atomic orbitals of Ti. We construct 3$p$, 3$d$ and 4$s$ like orbitals
using Wannier90 and use them as basis sets. The orbitals are shown in fig. 3(a), (b),
(c) and (d). 
WOOP (fig. 3(e)) has non-zero value at Fermi level indicating the metallic nature of Ti. The bands
close to Fermi level originate mostly from partially filled 3$d$ orbitals of titanium.
Also the 4$s$ orbital mixes with 3$d$ and electrons are transferred from 4$s$ to 3$d$.
3$d-$ and 4$s-$ like MLWFs  have 2.88 and 1.35 electrons per Ti atom respectively.
The 3$p$ orbitals are mostly at deep in the valence band (not shown in fig. 3(e))
and have negligible contribution near Fermi level.
The total number of electrons turns out to be 19.7 which can be improved by 
including more number of bands in construction of Wannier orbitals. 
From our WOHP calculation we find that the interactions near Fermi level are predominantly
due to these partially filled $d-$orbitals.
Fig. 3(f) shows the WOHP of Ti. At Fermi level the interaction is of bonding kind 
while the anti-bonding interaction is predominant in conduction band manifold.

\subsection{Carbon Nanotube: A low dimensional system}

\begin{figure}[h]
\begin{center}
\includegraphics[scale=0.6]{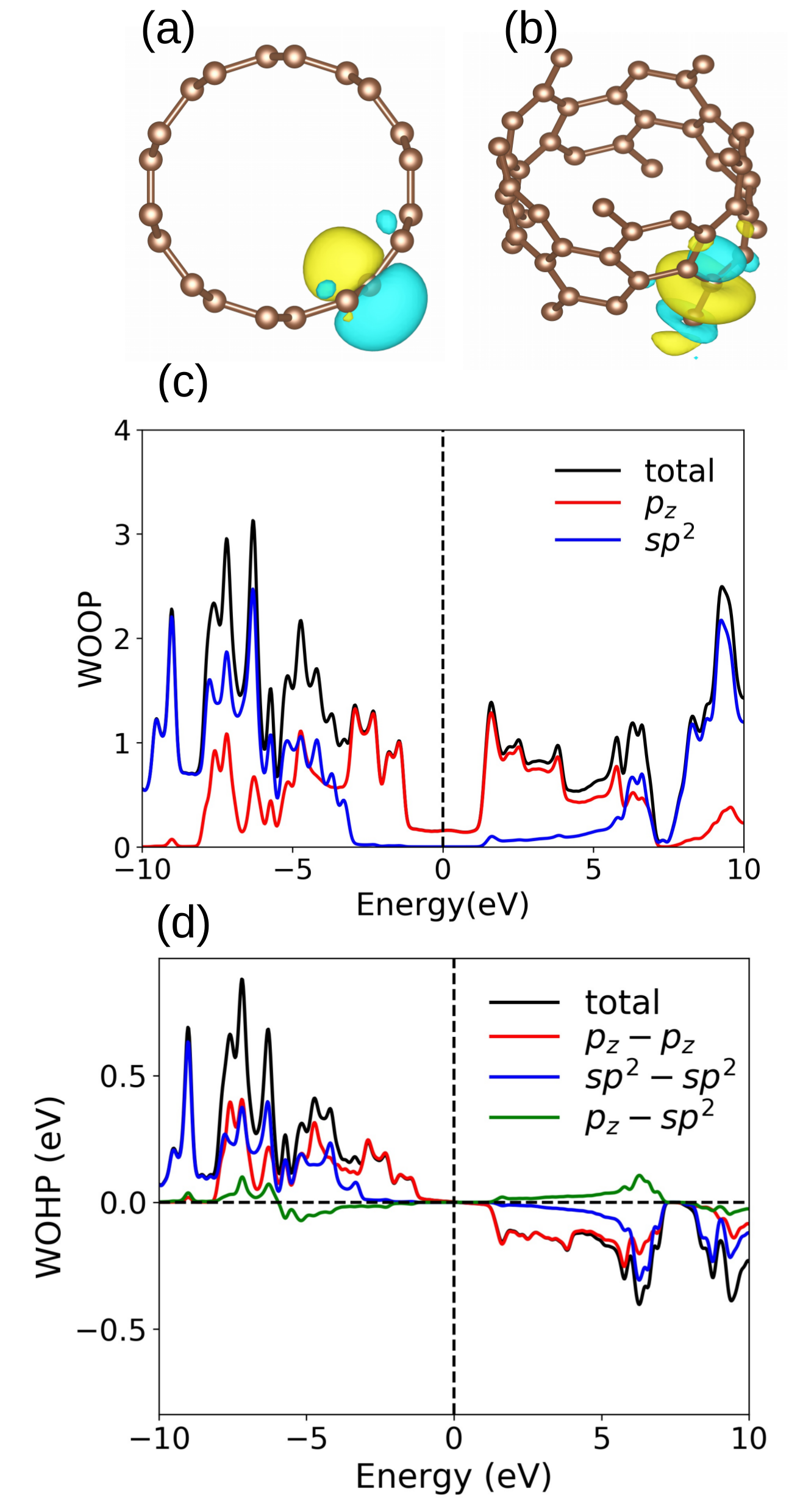}
\caption{(a) $p_z$ and (b) $sp^2$ like orbitals on C atom respectively. (c) shows the total
WOOP (black lines) and contribution from $p_z$ (red lines) and $sp^2$ (blue lines). (d) depicts WOHP and contribution of different interacting orbitals. Fermi level is set to zero.}
\end{center}
\end{figure}
Carbon nanotubes are an allotrope of carbon which can, depending on the configuration of the tube, either be a semiconductor or a metal.
Here we consider a (5,5) carbon nanotube (CNT(5,5)) which is a metal. 
CNT is periodic only along the axis of tube.
In CNT, the C 2$s$ and two 2$p$ orbitals hybridize to form $sp^2$ orbitals and $p_z$ has a lone electron. 
This example is different from the previous ones in the fact that
we construct $p_z$ and $sp^2$ orbitals at each atom following their local axes using Wannier90 instead of the
global axis of the tube along which the tube is oriented. The $p_z$ orbitals are directed
radially outward at each atom and the $sp^2$ orbitals, which are in the tangential plane, participate in bond formation. The orbitals are shown in
fig. 4(a) and (b). We calculate WOOP and WOHP using these orbitals. The black line in fig. 4(c)
show the total WOOP and the red and blue lines represent the contribution from $p_z$ and $sp^2$
respectively. From the fig. 4(c) it is evident that only $p_z$ contributes near the Fermi level
while $sp^2$ orbitals are deep in the valence band and conduction band continuum.
The $sp^2$ orbitals take part into bond formation and hold 59.9 electrons while the $p_z$ orbitals
contribute 19.9 electrons in the system. It is important to note that
if the orbitals were oriented following global axis, we would find contribution from $sp^2$ orbitals
near Fermi level. This would be qualitatively incorrect description of the system.
From the WOHP calculation (fig. 4(d)), it 
can be again seen that the $sp^2$ orbitals of two adjacent atoms result in bonding (anti-bonding) interactions below (above)
the Fermi level.
The $p_z$ orbitals also contribute to weak interaction near Fermi level and form $\pi$ bonding and $\pi^{*}$ anti-bonding interaction below and
above Fermi level respectively. On the other hand, interaction between $p_z$ and $sp^2$ orbital is negligible.
This is consistent with 
simple chemical analysis of the system.

\subsection{LNMO: A double perovskite}

\begin{figure}[h]
\begin{center}
\includegraphics[scale=0.5]{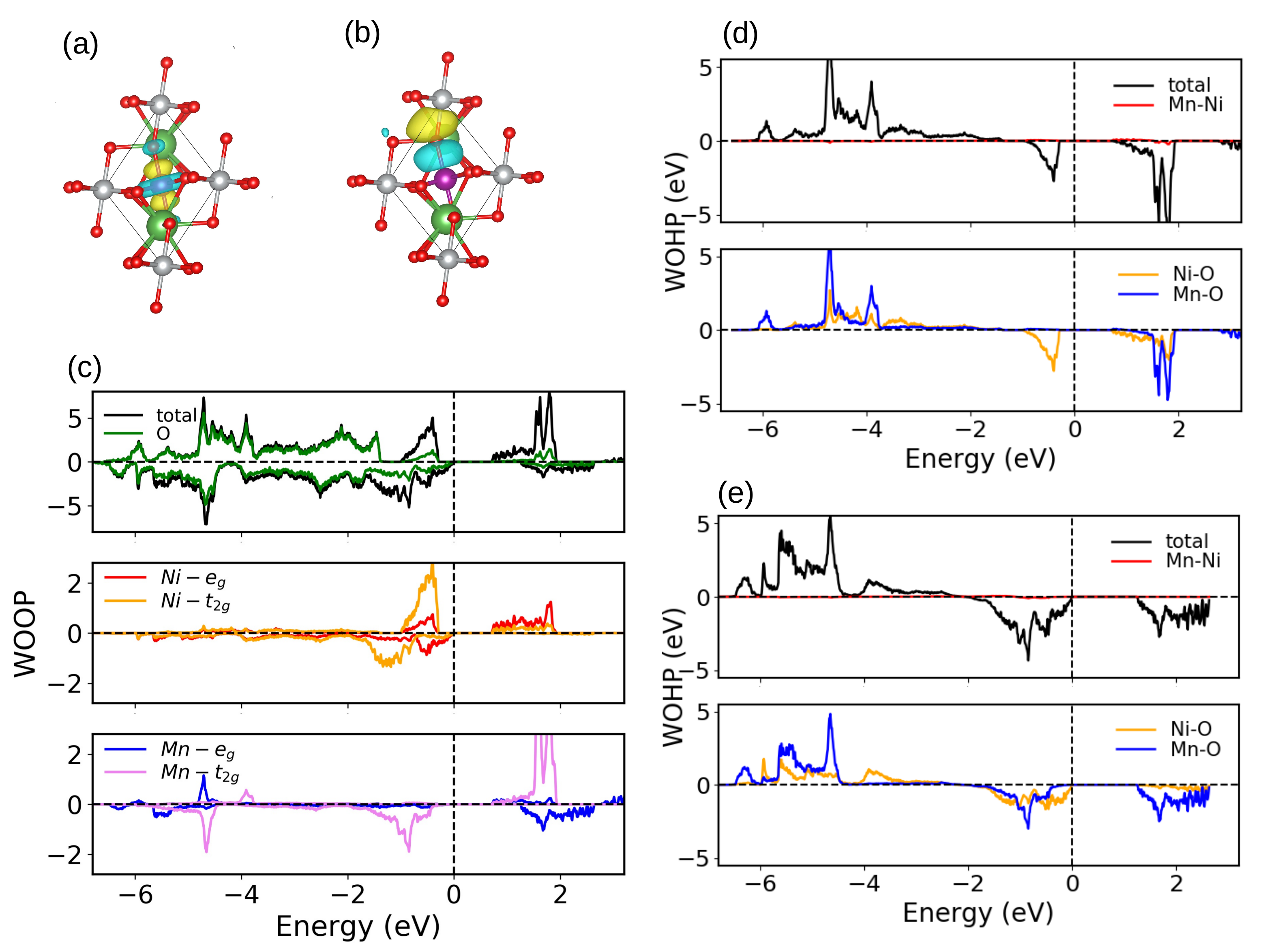}
\caption{(a) and (b) show $d_{z^2}$ at Mn (purple) and $p_z$ at O (red) atom respectively. La and Ni atoms are shown in green and grey colour respectively. (c) shows the WOOP  and orbital resolved
WOOP for up-spin (positive side) and down-spin (negative side). (d) and (e) depict the WOHP for up- and down-spin respectively. The valence band maximum of the system is set to zero.}
\end{center}
\end{figure}
Double perovskite La$_2$NiMnO$_6$ is a ferromagnetic insulator which can exist in either a rhombohedral or monoclinic phase.
LNMO has a distorted structure in which NiO$_6$ and MnO$_6$ octahedra are
tilted with respect to each other. Ni is in $2+$ state with electronic configuration d$^8$ ($t_{2g}^6e_{g}^2$)
while Mn$^{4+}$ has 3 electrons in its $d-$orbital (t$_{2g}^3$).
We investigate LNMO in rhombohedral phase. We perform a spin--polarized calculation. 
In LNMO the
$d-$orbitals are located at Ni and Mn atom sites. The oxygen $p-$orbitals mostly contribute to the states near the Fermi level.
Although there is some contribution from La 5$d$ orbitals, we focus on Ni$-4d$, Mn$-4d$ and O$-2p$
orbitals as they are responsible for the majority of the interactions. We orient the orbitals following the local
axis at each atom. In fig. 5(a) and 5(b), we show the constructed $d_{z^2}$ orbital at Mn and 
a $p-$orbital at oxygen respectively. To capture the correct contribution of the individual orbitals in electron
distribution or bond formation, it is important to orient the orbitals in right direction.
We construct the $d-$ and $p-$orbitals such that the local z-axis is along a bond direction.
With these constructed orbitals, we calculate WOOP 
for two spins as LNMO is ferromagnetic in rhombohedral phase.
Due to the crystal field splitting, we further group the orbitals in $t_{2g}$ and $e_{g}$ levels.
Fig. 5(c) shows the WOOP calculation. For differentiating between two spins, we plot the
down-spin on negative y-axis. Due to our approach of orienting axis locally along the orbital direction,
we are able to correctly describe the origin of peaks in WOOP calculation. 
Both spin of Ni-$t_{2g}$ contribute near valence band edge and the $e_{g}$ orbitals have smaller 
contribution. On the other hand only down-spin of Mn-$t_{2g}$ orbitals contributes near valence band edge.
This is consistent with the fact that Mn-$t_{2g}$ is half-filled but due to electron transfer from 
surrounding oxygen atoms, up-spin of Mn-$t_{2g}$ and Mn-$e_{g}$ have small contribution deep in the
valence band.
We calculate the number of electrons associated with $t_{2g}$ and $e_{g}$ orbitals for both Ni and Mn. 
We find that Ni has 3.4 ($t_{2g}$:2.5 and $e_{g}$:0.9) and 4.9 ($t_{2g}$:2.95 and $e_{g}$:1.95) spin--up
and down electrons respectively. Mn has 0.9 ($t_{2g}$:0.54 and $e_{g}$:0.44) and 3.9 ($t_{2g}$:3.0 and $e_{g}$:0.88)
electrons in spin--up and down channel respectively. The electron number differs from the occupancy of an
individual atom due to the electron transfer from surrounding atoms.
It is to be noted that in our formulation the orbitals in a group have same weight and 
the electrons are equally distributed among the orbitals in each group.
For example, each of the $t_{2g}$ and $e_{g}$ orbitals in Mn holds 1 and 0.44 spin--down electron respectively.
In contrast when we follow a global axis to construct our basis set, the orbitals of same group have
different weights and as a result, the electrons are distributed unevenly among the orbitals; such as
two orbitals of $t_{2g}$ set in Mn have 0.68 spin--down electron while the other one has 0.82.
This distribution is not correct as all the $t_{2g}$ orbitals are equivalent and should have same number of electron.
Furthermore, the $d_{z^2}$ of $e_{g}$ has 1 spin--down electron and $d_{x^2-y^2}$ has 0.82.
This discrepancy is because the orientation following the global axis breaks the equivalence of the orbitals.  
From our WOOP calculations we find that the number of spin-down electrons from both Ni and Mn is higher than spin-up  by 1.5 and
3 respectively which is consistent with the experimentally observed magnetic moments
of 1.9 $\mu_{B}$ and 3.0$\mu_{B}$ on Ni and Mn \cite{lmno_expt} and previous theoretical
calculations \cite{lmno2,lmno3}.
 Now we plot the WOHP for both spin--up and down in 
fig. 5(d) and 5(e) respectively. We find that there is no direct interaction between Mn and Ni.
Both Mn and Ni interact with oxygen strongly and this gives rise to the superexchange interaction between the 
$d-$orbitals of Ni and Mn mediated through oxygen. 
While the deep valence states give rise to bonding interactions in LNMO,
the Mn-O and Ni-O interactions are anti-bonding near band edges.

\section{Conclusion}
We have proposed and implemented an alternative formalism for population analysis using Wannier function as basis.
The uniqueness of this basis is that the spilling factor is zero for insulators or for the materials where the Wannier functions are constructed
from an isolated set of bands. For other case, where one has entangled bands, the spilling factor is low and it
can be further improved by better representation of the constructed Wannier
functions. We have performed five examples
using our formalism. Our diamond, GaAs and Ti examples match well with previous calculations. 
For the calculations on CNT and LNMO, the orbitals are oriented following local axes  which
accounts for the correct electron distribution among the orbitals.


\begin{acknowledgement}
We thank the Supercomputer Education and Research Centre (SERC) at IISc
for providing the computational resources. 
\end{acknowledgement}

\providecommand{\latin}[1]{#1}
\makeatletter
\providecommand{\doi}
  {\begingroup\let\do\@makeother\dospecials
  \catcode`\{=1 \catcode`\}=2 \doi@aux}
\providecommand{\doi@aux}[1]{\endgroup\texttt{#1}}
\makeatother
\providecommand*\mcitethebibliography{\thebibliography}
\csname @ifundefined\endcsname{endmcitethebibliography}
  {\let\endmcitethebibliography\endthebibliography}{}

\end{document}